\documentclass[Conference]{IEEEtran}
\usepackage{multirow}
\usepackage{multicol}
\usepackage{lipsum}
\usepackage{graphicx}
\usepackage{graphics}
\usepackage{amsmath}
\usepackage{amsbsy}
\usepackage{blindtext}
\usepackage{tcolorbox}
\usepackage{amssymb}
\usepackage{scalerel}
\usepackage{mathabx}
\usepackage{verbatim}
\usepackage{booktabs}
\usepackage{rotating,tabularx}
\usepackage{mathrsfs} 
\usepackage{hyperref}

\usepackage{adjustbox}
\usepackage{soul}
\usepackage{xcolor}
\usepackage{cite}
\usepackage{tikz}
\usepackage{color, colortbl}
\usepackage[first=0,last=9]{lcg}
\definecolor{LightGray}{rgb}{0.7,0.7,0.7}

\newcommand*{\rn}{\textcolor{black}}

\usepackage[wby]{callouts}
\usetikzlibrary{shapes.multipart}

\usepackage{array}
\usepackage{makecell}

\allowdisplaybreaks

\usepackage{amsthm}
\theoremstyle{definition}

\theoremstyle{remark}

\usepackage[utf8]{inputenc}
\usepackage[english]{babel}

\usepackage[caption=false,font=footnotesize]{subfig}
\usepackage[T1]{fontenc}
\usepackage{scalerel,stackengine}
\newcommand\reallywidecheck[1]{%
\savestack{\tmpbox}{\stretchto{%
  \scaleto{%
    \scalerel*[\widthof{\ensuremath{#1}}]{\kern-.6pt\bigwedge\kern-.6pt}%
    {\rule[-\textheight/2]{1ex}{\textheight}}
  }{\textheight}%
}{0.5ex}}%
\stackon[1pt]{#1}{\scalebox{-1}{\tmpbox}}%
}
\stackMath
\IEEEoverridecommandlockouts
\IEEEoverridecommandlockouts

\newif\ifarxiv
\arxivtrue
\arxivfalse

\newcolumntype{s}{>{} p{3.0cm}}
\newcolumntype{b}{>{} p{4.0cm}}

\renewcommand{\baselinestretch}{.90}

\usepackage[margin=1in,footskip=0.4in]{geometry}


\begin{document}

\title{\LARGE\bf
An Automated Framework for Assessing Electric Vehicle Charging Impacts on a Campus Distribution Grid}

\author{Mohammadreza Iranpour, Sammy Hamed, Mohammad Rasoul Narimani, Silvia Carpitella, Kourosh Sedghisigarchi, Xudong Jia
\thanks{Department of Electrical and Computer Engineering, California State University Northridge (CSUN). Rasoul.narimani@csun.edu. Support from NSF contract \#2308498 and the Climate Action - Community-driven eLectric vEhicle
chArging solutioN (CA-CLEAN) project}%
}

\maketitle

\begin{abstract}

\rn{This paper introduces a unified and automated framework designed to dynamically assess the impact of electric vehicle (EV) charging on distribution feeders and transformers at California State University, Northridge (CSUN). As EV adoption accelerates, the resulting increase in charging demand imposes additional stress on local power distribution systems. Moreover, the evolving nature of EV load profiles throughout the day necessitates detailed temporal analysis to identify peak loading conditions, anticipate worst-case scenarios, and plan timely infrastructure upgrades. Our main contribution is the development of a flexible testbed that integrates Julia, a high-performance programming language for technical computing, with PowerWorld Simulator via the EasySimauto.jl package. This integration enables seamless modeling, simulation, and analysis of EV charging load profiles and their implications for campus grid infrastructure. The framework leverages a real-world dataset collected from CSUN’s EV charging stations, consisting of 15-minute interval measurements over the course of one year. By coupling high-resolution data with dynamic simulations, the proposed system offers a valuable tool for evaluating transformer loading, feeder utilization, and overall system stress. The results support data-driven decision-making for EV infrastructure deployment, load forecasting, and energy management strategies. In addition, the framework allows for scenario-based studies to explore the impact of future increases in EV penetration or changes in charging behavior. Its modular architecture also makes it adaptable to other campus or urban distribution systems facing similar electrification challenges.}

\end{abstract}


\section{Introduction}
\label{Introduction}

\rn{The growing use of electric vehicles (EVs) is changing both transportation systems and how electric power is consumed and managed. This growth is supported by several factors, such as the decreasing cost of batteries, increasing awareness of environmental issues, and strong government incentives \cite{IEA2023GlobalEV, Lutsey2018BatteryCosts, Bloomberg2019EV}. According to the International Energy Agency (IEA), the number of EVs on the road surpassed 26 million in 2022, which is over ten times the number in 2013 \cite{IEA2023GlobalEV}. As more EVs are deployed, the demand for electricity increases significantly, placing additional pressure on existing power grid infrastructure \cite{, narimani2015dynamic, niknam2012efficient, Sundstrom2012Impact, Liu2015EVImpact, narimani2015effect, narimani2017}. This growing demand requires new approaches for grid planning and load management to ensure reliable and efficient operation~\cite{narimani2019efficient, narimani2016reliability, azizivahed2017new, narimani2018optimal, asghari2019method, narimani2017energy, narimani2019demand, narimani2017multi}.}

\rn{Despite the positive impact of EVs on reducing greenhouse gas emissions and improving energy efficiency, they also have the potential to create new challenges for electric power systems. The extra demand caused by EV charging stations can lead to several operational problems, including overloaded feeders, voltage instability, and higher stress on transformers \cite{Clement2010Impact, Shao2013EVImpact}. Most existing power distribution systems were not originally built to handle these types of fast-changing and locally concentrated loads, which makes it important to evaluate their impact carefully \cite{Bessa2012EVForecast, Richardson2010Impact}. As the number of EVs continues to increase, there is a growing need to strengthen transmission and distribution networks to support the additional demand \cite{Cai2014EVCharging, Deb2018EVImpact}. Without proper planning and system upgrades, these stresses may lead to equipment degradation and lower overall grid reliability\cite{boyaci2022spatio, iranpour2025assessing}.}

\rn{Static analyses, which look at only a single snapshot of the power system, are no longer enough to understand the growing complexity and time-varying nature of EV charging behavior. Traditional power flow studies often use fixed or average demand values, which overlook the fact that EV charging patterns can change quickly and are often unpredictable. These simplified assumptions may lead to inaccurate evaluations of grid performance, especially as the number of EVs continues to grow. As EVs become a more common mode of transportation, the shortcomings of static methods become increasingly evident. Dynamic analysis methods are therefore necessary to model changes in load over time and to better assess potential operational issues, such as voltage drops, line congestion, and transformer overloading \cite{Yao2014Dynamic, Wang2020DynamicAssessment}. Incorporating time-dependent simulations into planning and operations helps utilities anticipate critical stress points and develop more effective mitigation strategies.}

\rn{The changing nature of EV load profiles is influenced by several connected factors, such as time-of-use electricity pricing, individual charging habits, differences in vehicle models, the availability of charging stations, and battery charging strategies. For instance, many users may plug in their vehicles at the same time, such as after work or after classes, leading to short-term spikes in demand that go well beyond typical average levels. These unpredictable patterns make it difficult to depend on simple forecasting tools or fixed models. As a result, real-time or near-real-time simulations that use detailed, high-frequency data are essential for accurately analyzing how EVs affect the power grid under realistic conditions \cite{Mohamed2015EVModeling, Tan2016LoadProfile}. This level of analysis is particularly important for system planners who must ensure grid reliability while managing increasingly flexible and variable energy demands.}

\rn{University campuses, such as California State University Northridge (CSUN), are well-suited for studying the real-world effects of EV charging on power distribution systems. Campuses typically have a high concentration of buildings and a mix of different types of electricity use, including residential, academic, and research loads. They also serve a wide range of EV users, faculty, staff, and students' vehicles. As more EV charging stations are installed, managing when and where charging occurs becomes more complex, both over time and across different campus locations. These conditions make campuses ideal environments for testing how clustered charging patterns, driven by daily schedules and user habits, can lead to overloading of transformers and congestion on local feeders \cite{Gough2019CampusMicrogrid, Wu2018CampusEV}.}

\rn{To accurately study complex and data-rich scenarios like EV charging, it is important to use computational tools that are both powerful and flexible. Interface-based simulation platforms have become effective solutions for this purpose. These tools make it possible to connect high-performance programming languages, such as Julia, with professional power system simulation software like PowerWorld. This setup allows users to automate simulation tasks, adjust inputs such as EV charging patterns or building loads, and run real-time power flow analysis across many different scenarios \cite{Lee2016SimInterface, Fletcher2019PowerSim}. These interfaces improve the scalability, consistency, and accuracy of power system studies, helping researchers explore how the grid responds under a wide range of operating conditions \cite{Huang2020Framework, Martinez2021DynamicInterface}. When working with large datasets, such as the one-year set of 15-minute interval measurements from CSUN’s EV charging stations, these frameworks are especially important. They offer a reliable and adaptable environment for understanding how EV charging affects the grid over time and for planning future infrastructure upgrades with greater confidence.}

\begin{figure}
    \centering
\includegraphics[scale=0.23,trim=8.5cm 0.3cm 6cm 0.5cm,clip]{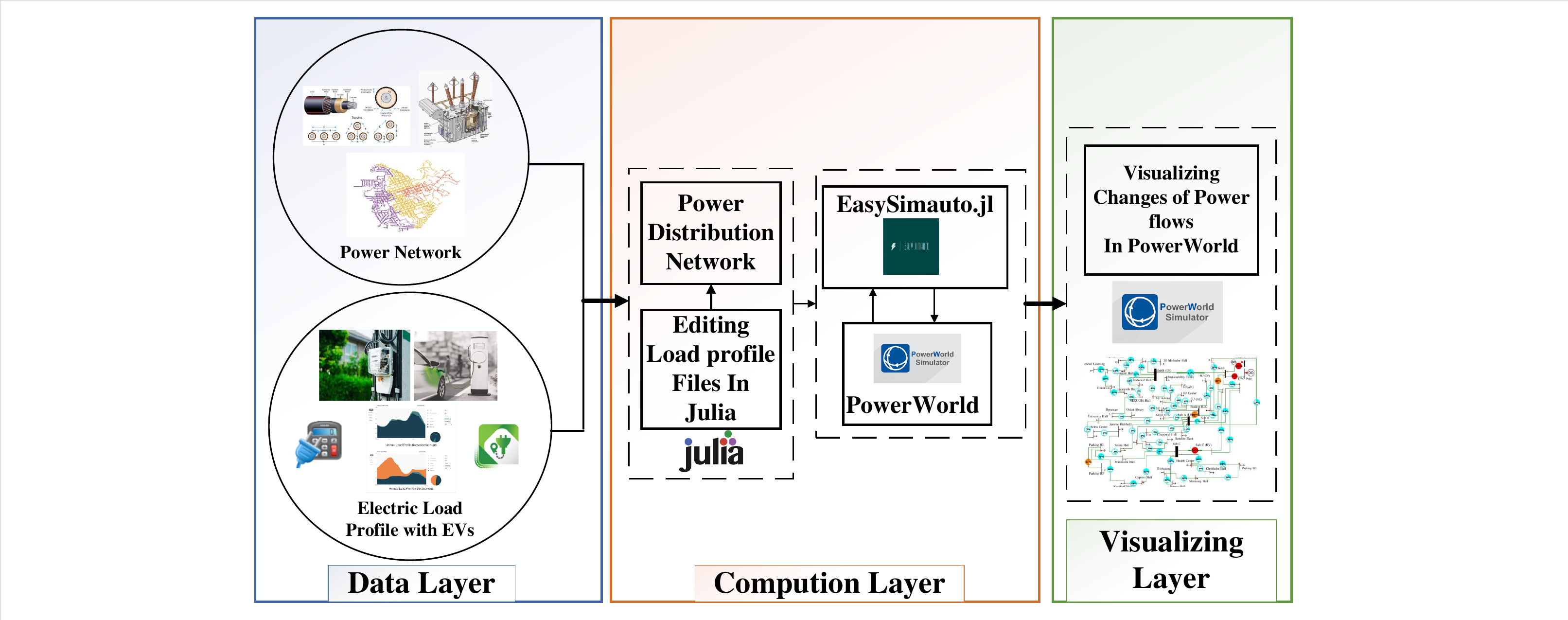}
    \caption{Schematic of the proposed framework for Dynamically Investigating the Impact of EV Chargers on Power Distribution Systems using Various tools and in three layers.}
    \label{fig:procedure}
\end{figure}

\rn{In this paper, we present an automated and unified framework to study how EV charging affects the electric power grid at CSUN. We created a detailed model of CSUN’s distribution network in PowerWorld, using real data about the campus infrastructure and layout. To support the analysis, we collected one year of electricity usage data from meters installed at campus buildings and EV charging stations, with measurements taken every 15 minutes. From this dataset, we identified the worst-case daily load profiles for analysis. Using the EasySimauto.jl package, we built a dynamic connection between Julia and PowerWorld to automate data input and simulation processes. We developed scripts that automatically adjusted the electrical load at different buses and ran power flow simulations for each load scenario. This approach allowed us to examine important system parameters such as voltage levels, transformer loading, and feeder congestion under different operating conditions. The main contribution of this work is the introduction of a flexible and automated framework that captures the time-varying nature of EV-related load profiles and allows system parameters to be modified at any selected time interval. This capability enables the simulation of a wide range of operating scenarios and provides a powerful tool for both power system operation and planning. The framework supports detailed, time-resolved analysis of grid impacts, making it highly valuable for infrastructure design, reliability assessment, and long-term energy management.
}

\rn{The rest of this paper is organized as follows:
In Section \ref{sec:Methodological Framework}, we describe how CSUN’s campus electric grid is modeled and coresponding data sets are injected to the proposed framework. In Section \ref{sec:highmodel} we introduce the proposed
high-fidelity model for analyzing the impact of EVs on
low-voltage distribution systems. Section \ref{sec:results} presents the
simulation results, and Section \ref{sec:highmodel} provides the conclusion.}

\section{Methodological Framework}
\label{sec:Methodological Framework}

\subsection{Modeling CSUN’s Electric Power Network}

\begin{figure}
    \centering
\includegraphics[scale=0.325,trim=0cm 0.0cm 0cm 0.0cm,clip]{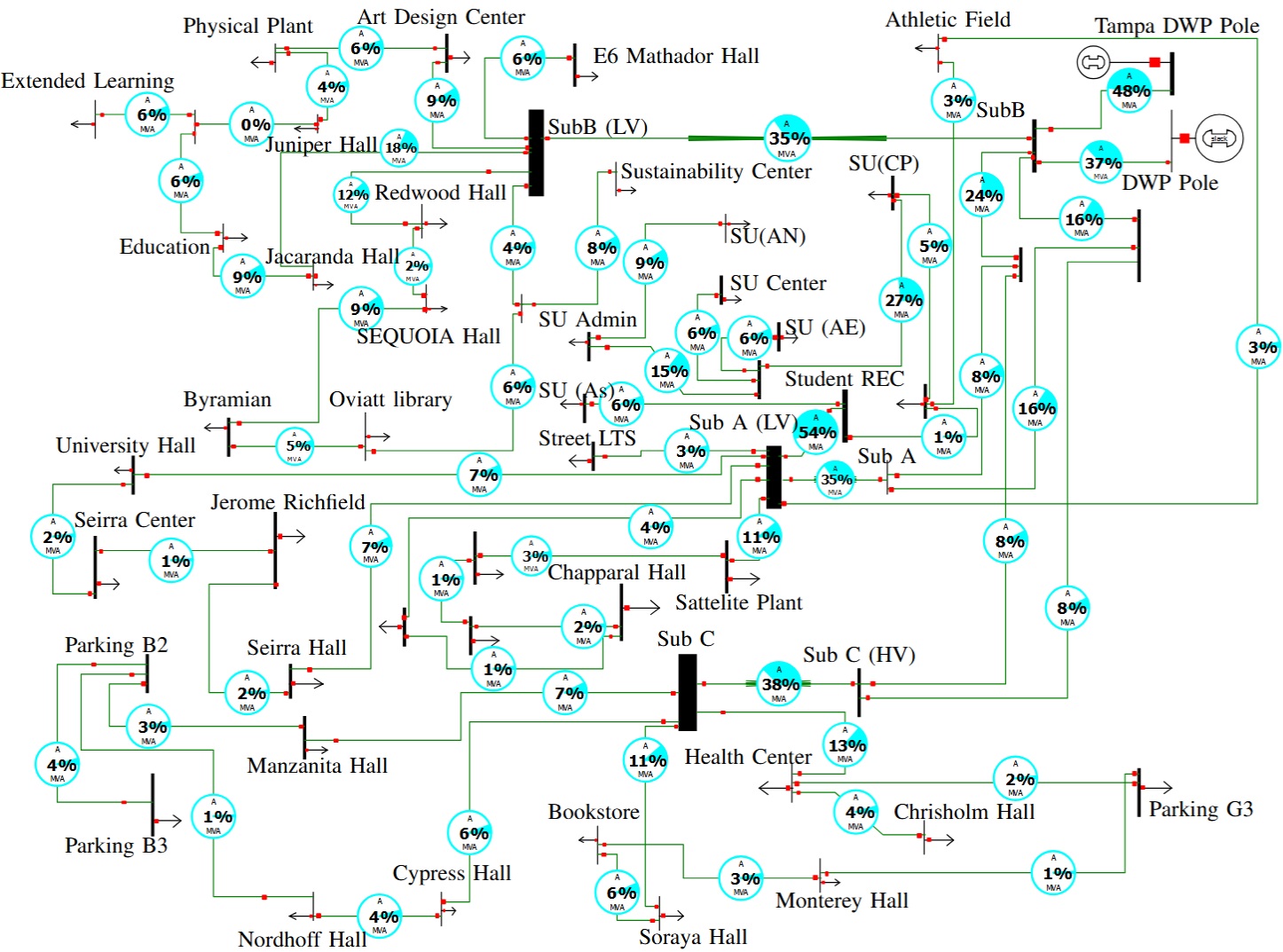}
    \caption{PowerWorld model of CSUN’s campus distribution network under regular (non-EV) load conditions. All feeders operate below thermal limits, indicating no line congestion in the baseline scenario.}
    \label{fig:csun_grid}
\end{figure}

\rn{To accurately assess how EV charging affects grid operations, it is important to have a realistic model of CSUN’s electric power distribution system. An overview of the modeled campus distribution network in PowerWorld is shown in Figure~\ref{fig:csun_grid}. For this purpose, we built a detailed representation of the campus grid using data provided by the Physical Plant Management (PPM) department. The model includes key system components such as underground cables, transformers, and distributed generation units.} 

\rn{To calculate the resistance and impedance of the underground cables, we used a one-line diagram of the campus that showed the types of cables and their geographic layout. The coordinates of each cable were used to estimate their physical lengths. We then applied standard resistance values (in ohms per mile), provided by PPM for each cable type, and multiplied them by the cable lengths to determine total resistance. This process allowed us to assign realistic impedance values to each section of the network, preserving both spatial and physical accuracy in the simulation. Transformers were also modeled in detail using nameplate data from PPM, which included their capacities, impedances, and configurations. These details were incorporated into our PowerWorld simulation to better understand how transformers respond to changes in load, particularly under EV charging conditions. In addition, the campus includes several photovoltaic (PV) systems that contribute to local energy generation. These include a 467 kW system at Parking Lot B2, a 225 kW system at Parking Lot E6, and a 1.2 MW system at the Student Recreation Center. These PV systems were modeled as distributed generators using actual generation profiles provided by PPM, which allowed us to capture their variability over different times of the day and seasons.}

\subsection{Load Data Collection and Modeling}

\rn{To build a realistic and time-varying load model for the CSUN campus, we carried out a detailed data collection process that included measurements from both campus buildings and EV charging stations. Meters were installed on each building’s power cables to record real and reactive power demand every 15 minutes over the span of a full year. Since the campus electrical system operates close to a balanced condition, measuring just one phase per building was sufficient to represent the overall load. This approach simplified the data collection while still providing reliable results. We also analyzed the data to capture how electricity use changes over time, including daily cycles and seasonal trends. These time-based patterns are important for accurately reflecting real operating conditions. By incorporating this time-series data into our simulations, we were able to model how campus loads vary in response to factors like academic calendars, building occupancy, and routine operations. This time-resolved load modeling is essential for evaluating the true impact of EV integration and identifying periods of potential stress on the distribution system.
}

\subsection{Incorporating EV Charging Loads}

\rn{Including EV charging loads in the campus power system model was a key part of our analysis. We used real data collected from CSUN’s operational EV charging stations. This dataset contained one year of charging records at 15-minute intervals for each station, providing detailed information on how much energy was used, how often vehicles were charged, and when charging typically occurred.}

\rn{To study the effects of this additional load, we identified the times with the highest charging activity and created daily worst-case charging profiles. These were combined with the building-level load data to produce a complete load profile that represents the campus system under heavy demand. Our model included both real power (kW) and reactive power (kVar) components of EV charging to accurately reflect their effect on the power factor. By directly embedding these EV charging patterns into our time-based load profiles, we developed a strong and realistic framework for evaluating how EVs impact system performance, specifically in terms of voltage fluctuations, transformer stress, and feeder congestion, within the current campus grid setup under various operating scenarios.}

\subsection{Power Flow Analysis via PowerWorld Simulator}

\rn{To study how the campus grid responds to changing load conditions and EV charging activity, we used PowerWorld Simulator, an industry-standard tool for performing time-series power flow analysis. We connected PowerWorld with Julia using the EasySimauto.jl interface, which allowed us to automate simulations and manage data efficiently.}
\rn{This setup made it possible to update the load at each bus in the system dynamically and run power flow calculations across thousands of time steps, each representing a different loading condition. Our Julia scripts processed the time-series data from building and EV loads and sent the updated values directly to PowerWorld using SimAuto commands. This automated approach eliminated the need for manual data entry or interface interaction, which greatly reduced the overall simulation time and made the process more consistent and repeatable.  In addition, the flexible scripting environment allowed us to easily test different load growth scenarios and EV penetration levels. This capability makes the framework suitable for long-term planning studies and stress testing of grid infrastructure under future conditions.}

\section{High-Fidelity Model}
\label{sec:highmodel}

\rn{In this study, we developed a detailed and structured framework to evaluate how residential EV chargers affect low-voltage power distribution systems. The framework is organized into three main layers: the data layer, the computational layer, and the visualization layer. These layers work together to process input data, run simulations, and present the results clearly. An overview of the complete framework is shown in Fig.~\ref{fig:procedure}.}

\begin{figure}[ht]
    \centering
    \includegraphics[width=0.95\linewidth]{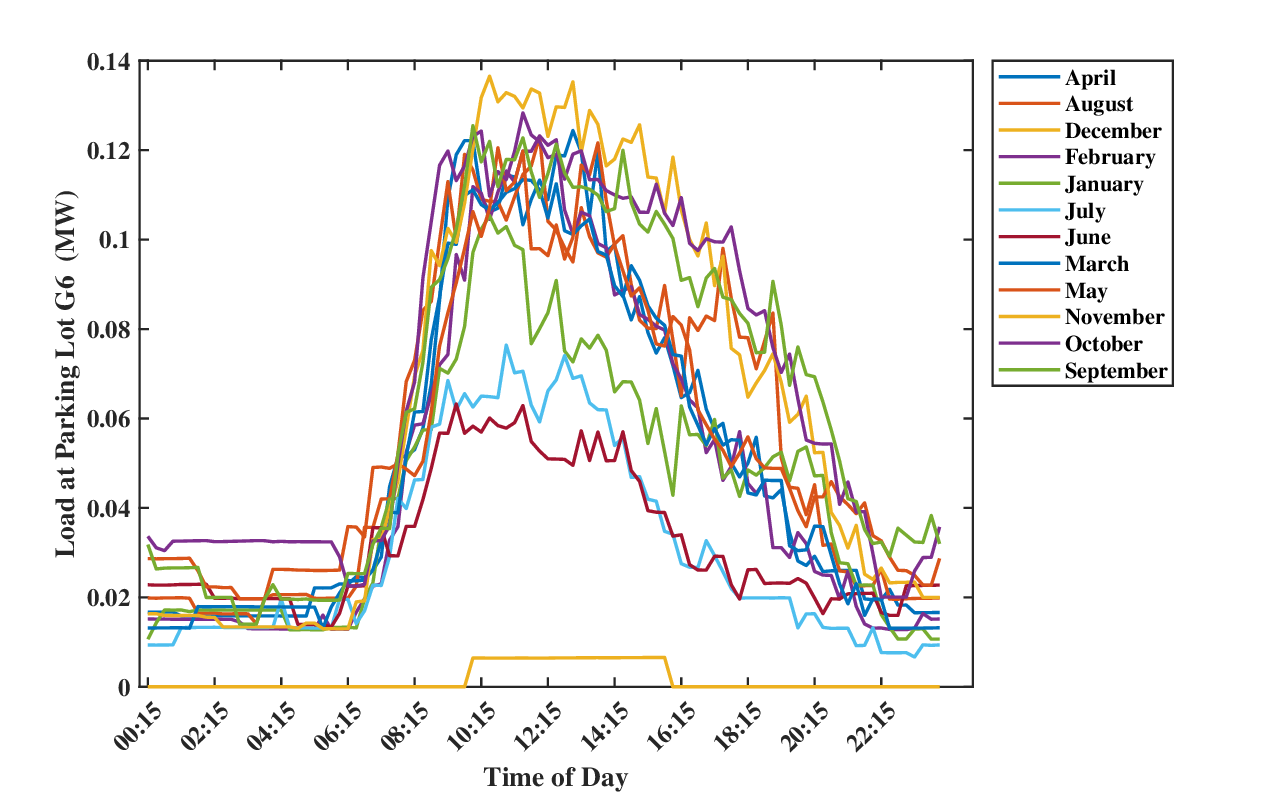}
    \caption{Load profiles of EV charging station at Parkin Lot G6 during one day in 12 months.}
    \label{fig:load profile}
\end{figure}

In the Computational Layer, each of these load profiles was applied to the simulated distribution network using the EasySimAuto.jl API. This allowed for programmatic updates to PowerWorld’s load values for each interval, automating the process of time-series simulation and ensuring consistent system modeling across all months. Each scenario was simulated in PowerWorld, and the resulting voltage magnitudes, line power flow's changes, and line losses levels were recorded for analysis.

\subsection{Data Layer}

\rn{The data layer is responsible for gathering and organizing all necessary input data for the simulation process. It consists of two main components: First, we collected detailed information about the physical characteristics of the campus distribution network, including specifications for underground cables, transformers, and the overall system topology. This information was used to build a realistic and accurate network model, which was implemented in PowerWorld Simulator. Second, the data layer incorporates time-series measurements for both the baseline building loads and the EV charging profiles. As described in the previous section, these data were recorded at 15-minute intervals over a full calendar year. For each month, we analyzed the data to identify the day with the highest peak load, representing the most critical or worst-case scenario. This process resulted in twelve representative daily load profiles, one for each month, each consisting of 96 time steps to reflect the 15-minute resolution.}

\subsection{Computational Layer}

\rn{The computational layer handles the automated execution of simulations using PowerWorld Simulator. We used the EasySimAuto.jl package, a Julia-based interface, to control and interact with PowerWorld programmatically. This setup allowed us to load different load profiles into the simulation model automatically, without manual input.}

\rn{EasySimAuto.jl works by connecting to PowerWorld’s SimAuto server, which provides access to a range of commands and data through its COM-based interface. Using this API, we updated load values for each 15-minute interval in the selected daily profiles and then ran simulations to observe the system's behavior under various EV charging scenarios. This layer made it possible to run flexible and repeatable experiments by automating the setup, execution, and data retrieval processes for each simulation case.}

\subsection{Visualization Layer}

\rn{The visualization layer is responsible for analyzing and presenting the results from the simulations based on the different load profiles and time intervals. From each PowerWorld simulation, we extracted key performance indicators such as voltage levels, current flow, and loading on network components. These results were post-processed and visualized using custom scripts written in Julia. The goal was to understand how different EV charging behaviors, captured for the most critical day of each month, affect the performance and stability of the campus power system.}

\begin{figure}[h]
  \centering
  \includegraphics[width=.80\linewidth]{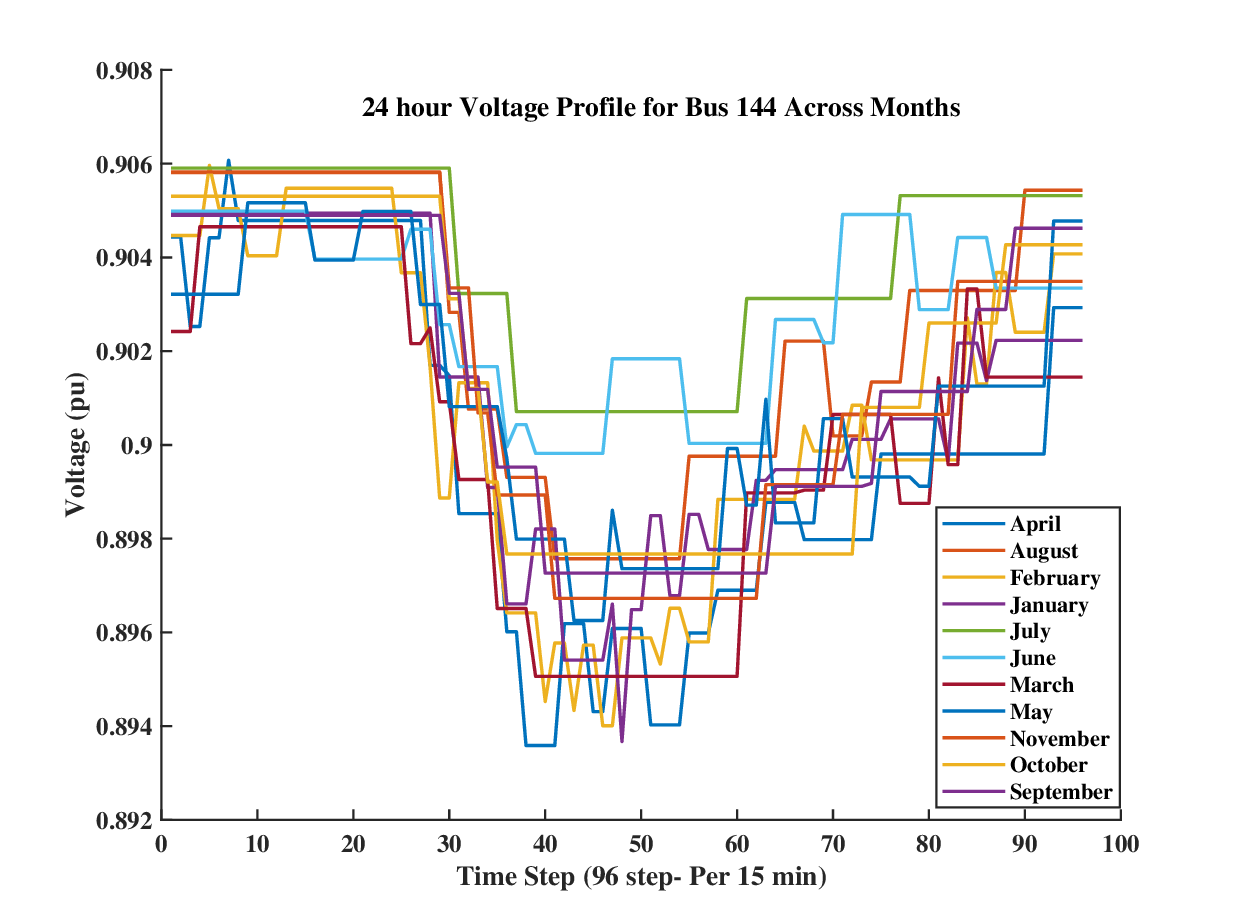}
  \caption{24-hour per-unit voltage profiles at Bus 144 (Parking Lot G6) for each month under the EV load scenario.}
  \label{fig:bus144_profiles}
\end{figure}

\begin{figure*}
  \centering
  \includegraphics[scale=.65,trim=0.2cm 0.5cm 0.5cm 0.05cm,clip]{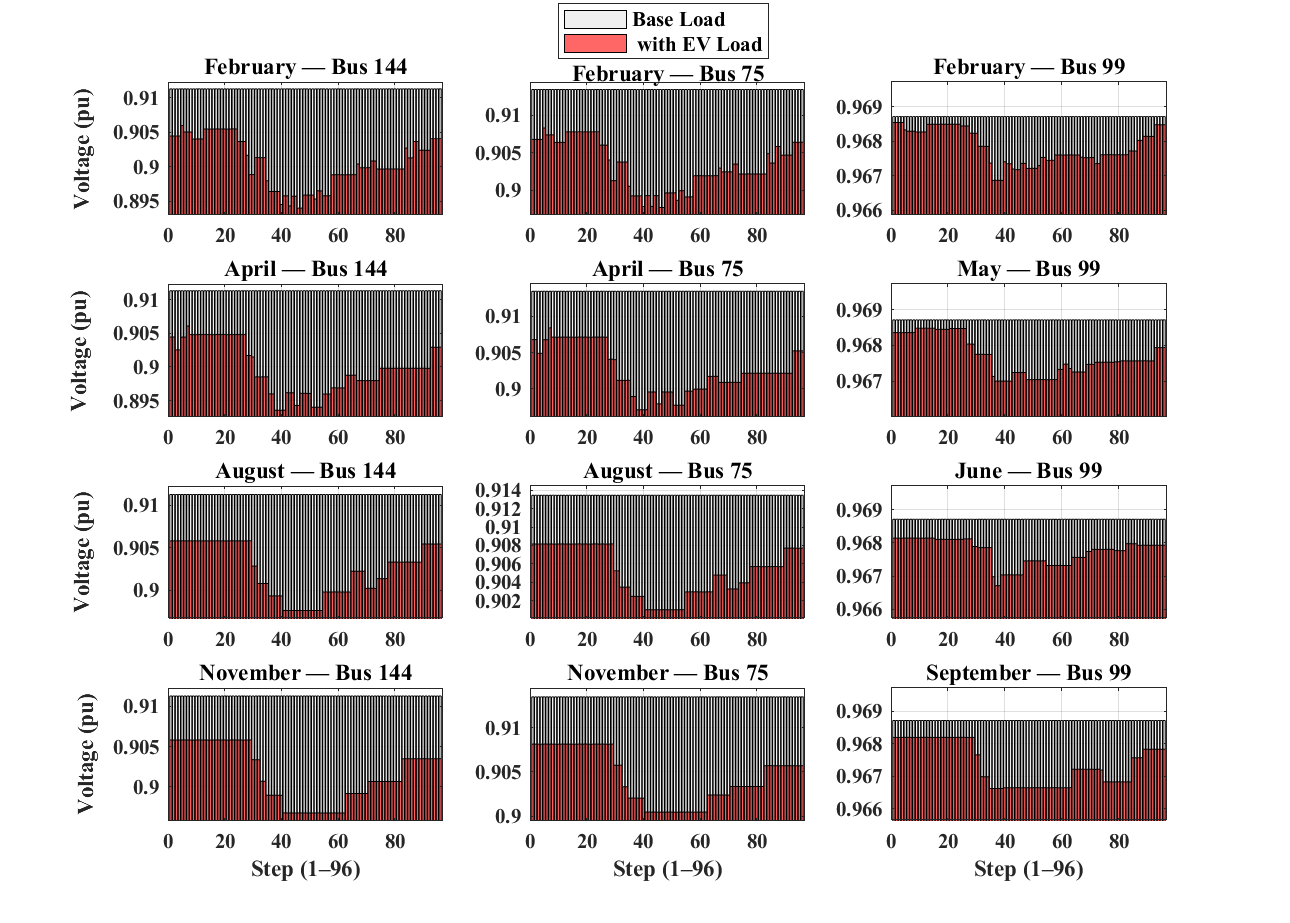}
  \caption{%
    Twenty-four-hour voltage trajectories at three critical CSUN grid under both the base load and a 4× EV-load scenario. Each subplot overlays the base-case voltage (gray bars) with the EV-augmented voltage (red bars) over 96 fifteen-minute steps. For each bus, 144, 75 and 99, the single worst-swing month (February, April, August or November) was automatically selected based on the maximum difference in voltage range when EV load is multiplied by four.
  }
  \label{fig:voltage_profiles}
\end{figure*}

\rn{For each of the twelve monthly profiles (with 96 time intervals per day), we tracked system responses such as voltage drops, overloaded lines, and transformer stress levels. These results were plotted over time to identify when and where the grid experiences the most stress. In addition, we created comparative heatmaps and time-series plots to show both the spatial and temporal effects of residential EV charging on the distribution network. This layer provides a detailed, data-driven view of grid performance, helping utility planners and researchers identify critical time windows and system locations that may require upgrades or mitigation strategies.}

\section{Results and Discussion}
\label{sec:results}

\rn{To evaluate the impact of residential EV charging on low-voltage distribution systems, we ran simulations using twelve representative daily load profiles, one per month, based on year-long, 15-minute interval data. Each profile reflects the peak-load day of its month, capturing worst-case conditions in 96 time steps. These profiles represent realistic daily and seasonal variations across the campus. An example for Parking Lot G6 on a peak day is shown in Fig.~\ref{fig:load profile}.}

\rn{In the Visualization Layer, post-processed data were used to examine voltage trends, line flows, and power losses. Fig.~\ref{fig:bus144_profiles} shows the 24-hour per-unit voltage profile at Bus 144 (Parking Lot G6), which experiences the largest daily voltage swing under a heavy EV charging scenario (4× base EV load). Each trace represents one month’s 96 time steps at 15-minute intervals. A midday voltage plateau ($\approx$0.905 pu) drops sharply in the afternoon (steps 30–60), coinciding with peak charging demand. Spring months (e.g., February, April) show the lowest voltages ($\approx$0.893 pu), while summer months (June, July) remain slightly higher ($\approx$0.900-0.902 pu) due to lower connected EV during the summer. Voltage recovers after step 50 as charging activity declines. These results indicate that even with currently low EV penetration at CSUN, significant voltage drops already occur during peak hours. As EV adoption increases, these effects will intensify. Proactive mitigation, such as on-load tap changers or local energy storage, will be essential to maintain voltage stability, particularly at critical nodes like Parking Lot G6.}

\rn{Fig.~\ref{fig:voltage_profiles} shows the 24-hour voltage profiles at three critical CSUN buses, Bus 144 (Lot G6), Bus 75 (Lot B5), and Bus 99 (Lot G3), under both base load and a 4× EV load scenario. Each panel compares the voltage without EV charging (gray bars) to the voltage with EV load (red bars), across 96 fifteen-minute intervals. At Bus 144, the largest voltage drop occurs during February and April, reaching up to 0.015 pu below the base case. This indicates limited voltage support and moderate background load. Bus 75 shows a smaller but still noticeable drop of about 0.010 pu. Bus 99 is least affected, with voltage deviations generally below 0.005 pu, suggesting stronger upstream support. Seasonal variation is evident: voltage troughs are deeper and occur earlier in the winter and spring months, while summer and fall months (e.g., August and November) maintain higher minimum voltages due to stronger base conditions. These results demonstrate that concentrated EV charging can significantly degrade voltage quality at specific nodes. Even with current low EV adoption, the impact is measurable. As EV penetration increases, proactive measures, such as voltage regulation devices or infrastructure upgrades, will be necessary to maintain power quality and grid stability.}

\begin{figure}
  \centering
  \includegraphics[scale=.45,trim=0.2cm 0.5cm 0.5cm 0.1cm,clip]{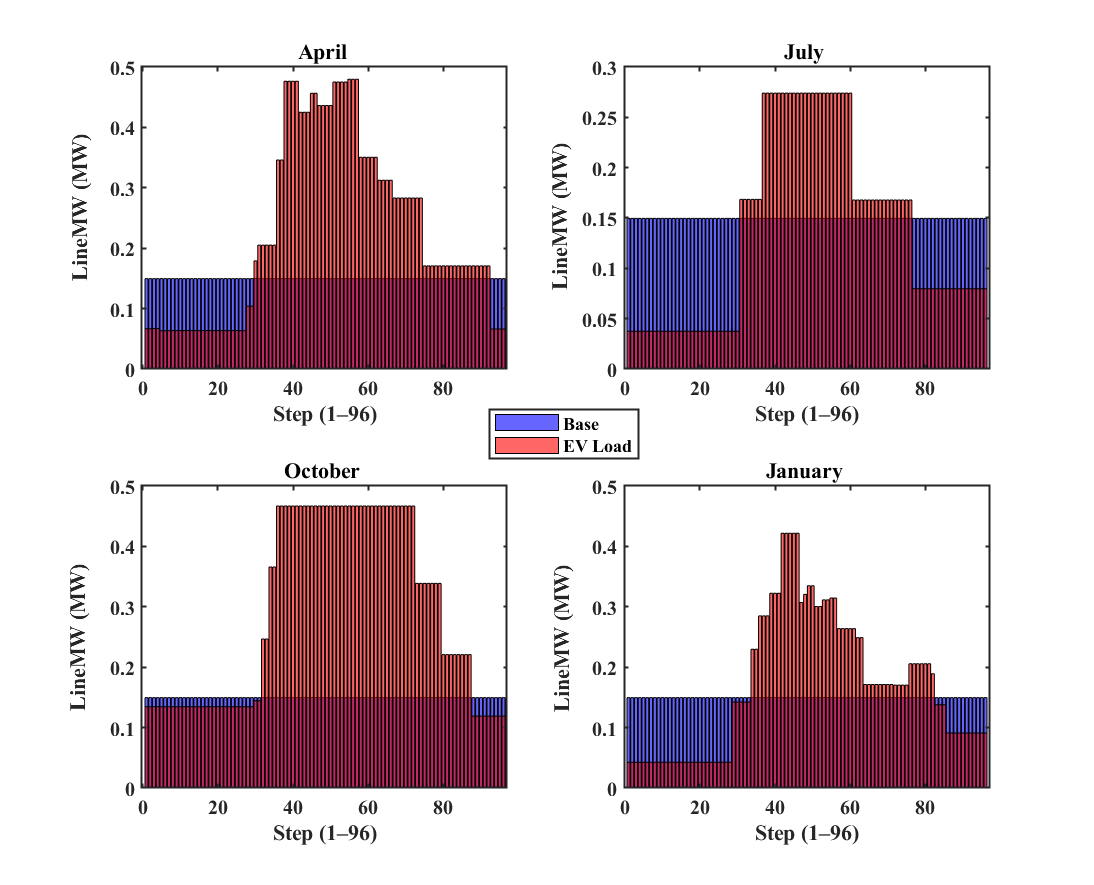}
  \caption{Comparison of 96‐step active power flows (LineMW) on the line between buses 144 (parking lot G6) and 145 under the base loading and EV charging scenarios for April, July, October, and January.}
  \label{fig:line144-145-flows}
\end{figure}

\rn{Figure~\ref{fig:line144-145-flows} compares active power flows between Buses 144 (Lot G6) and 145 over 96 time steps for four representative months: April, July, October, and January. Under base load conditions (blue bars), the power flow remains steady at around 0.15~MW, reflecting a relatively constant campus load. When EV charging is included (red bars), line loading increases significantly during the day. Peak flows occur between steps 35-65, reaching up to 0.48~MW in April and October, 0.42~MW in January, and 0.27~MW in July. The higher peaks in April and October align with increased EV usage in moderate weather, while the lower peak in July reflects reduced background load. These results show that uncoordinated EV charging can more than triple the load on specific feeders, potentially violating thermal or voltage limits. The step-by-step analysis highlights the need for demand management or infrastructure upgrades to accommodate future EV growth on the CSUN campus.}



\section{Conclusion}

\rn{This paper presents an automated, high-resolution framework for assessing the impact of EV charging on the CSUN distribution network. By integrating real 15-minute EV charging data with Julia’s EasySimauto.jl interface and PowerWorld Simulator, we developed a fully scriptable pipeline that performs time-series load flow simulations and extracts voltage and power flow metrics across the network. Our analysis highlights significant daily and seasonal voltage drops and feeder loading increases at key campus locations, even under current low EV penetration. These effects are most pronounced during afternoon peak charging periods, especially in spring and winter months. As EV adoption grows, such stresses are expected to intensify, potentially compromising power quality and system reliability. The proposed framework enables detailed, data-driven evaluation of feeder overload risks and voltage deviations under various load scenarios. This supports informed decision-making for coordinated charging strategies, infrastructure upgrades, and grid reinforcement. Built on a reproducible and scalable platform, this tool can assist utilities, planners, and researchers in developing and testing EV integration strategies under realistic conditions.}

\bibliographystyle{IEEEtran}
\IEEEtriggeratref{40}
\bibliography{ref}
\end{document}